\documentclass[a4paper,12pt]{article}

\usepackage{amsmath,amsfonts,amssymb}
\usepackage{amsthm}
\usepackage{graphicx}
\usepackage{a4wide}
\usepackage{hyperref}

\numberwithin{equation}{section}

\newcommand{\Ne}[1]{\ensuremath{\mathcal{N}={#1}}} 
\newcommand{\ls}{\ensuremath{{l_s}}} 
\newcommand{\gs}{\ensuremath{{g_s}}} 
\newcommand{\ket}[1]{\lvert #1 \rangle} 
\newcommand{\hyp}[2]{\ensuremath{(\mathbf{#1},\mathbf{#2})}} 
\newcommand{\gym}{{\ensuremath{g_{\text{YM}}^2}}} 
\newcommand{\abs}[1]{{\ensuremath{\left\lvert #1 \right\rvert}}} 
\newcommand{\hd}[1]{\ensuremath{\phantom{ }^{\star_{#1}}}} 

\newcommand{\cZZ}[1]{{\ensuremath{\mathbb{C}^3/\mathbb{Z}_{#1}\times\mathbb{Z}_{#1}}}} 
\newcommand{\czz}{\cZZ{2}} 

\begin{document}

\begin{titlepage}

\begin{flushright}
{\small
DFTT 20/2003\\
hep-th/0310157\\
}
\end{flushright}

\begin{center}

\vspace{1.5cm}

{\huge Non-Perturbative Gauge}

\vspace{0.2cm}

{\huge Superpotentials from Supergravity}

\vspace{1.5cm}

{\large {\bf Emiliano Imeroni $^{a,c}$}\,\, and\quad {\bf Alberto Lerda $^{b,a,c}$}}

\vspace{1cm}

$^a$ \emph{Dipartimento di Fisica Teorica, Universit\`a di Torino,\\
Via P. Giuria 1, I-10125 Torino, Italy}

\vspace{0.2cm}

$^b$ \emph{Dipartimento di Scienze e Tecnologie Avanzate\\
Universit\`a del Piemonte Orientale, I-15100 Alessandria, Italy}

\vspace{0.2cm}

$^c$ \emph{I.N.F.N., Sezione di Torino, Via P. Giuria 1, I-10125 Torino, Italy}

\vspace{0.8cm}
email: \quad\verb|imeroni@to.infn.it|\,,\quad\verb|lerda@to.infn.it|\\

\end{center}

\vspace{2cm}

\begin{abstract}
We study $U(N)$ SQCD with $N_f\leq N$ flavors of quarks and antiquarks by engineering it with a configuration of fractional D3-branes on a \czz{} orbifold. In particular we show how the moduli space of the gauge theory naturally emerges from the classical geometry produced by the D3-branes, and how the non-perturbatively generated superpotential is recovered from geometrical data.
\end{abstract}

\end{titlepage}

\tableofcontents

\vspace{0.5cm}

\section{Introduction}\label{s:intro}

The search for extensions of the gauge/gravity correspondence 
towards non-conformal theories with a reduced number of supercharges
has recently achieved some interesting progress.
Indeed, it has been shown that many relevant properties 
of non-conformal \Ne{1} and \Ne{2} supersymmetric gauge theories 
in four dimensions can be extracted from dual supergravity solutions 
associated to different set-ups of strings and D-branes.
For example, at the perturbative level one can get the correct 
logarithmic running of the coupling constant and 
the chiral anomaly, while at the non-perturbative level one 
can discuss instantons, gaugino condensation and confining strings.

Many of these results can be obtained by studying in detail the supergravity
solutions produced by stacks of fractional D3-branes in conifold~\cite{Klebanov:1999rd,Klebanov:2000nc,Klebanov:2000hb,Loewy:2001pq,
Imeroni:2002me} and 
orbifold~\cite{Klebanov:1999rd,orbifold,
Bertolini:2001gg,Marotta:2002gc} backgrounds. 
Alternatively, non-conformal gauge theories with
\Ne{1} or \Ne{2} supersymmetry can be realized by wrapping D-branes on suitable
supersymmetric cycles in Calabi--Yau or K3 
manifolds~\cite{MN}. Also in these cases, 
a detailed study of the corresponding supergravity solutions has provided 
relevant information on the dual gauge theory
~\cite{Loewy:2001pq,wrapped}. Most of these developments are 
covered in recent reviews~\cite{Herzog:2002ih,reviews}
to which we refer also for a more extended bibliography.

Another line of research has been the study of the so-called
``geometric transitions''~\cite{Gopakumar:1998ki,Vafa:2000wi}, 
where one engineers gauge theories by 
wrapping D5-branes on two-cycles of resolved Calabi--Yau manifolds 
in such a way that the geometry flows to a deformed manifold 
where branes are replaced by fluxes through the newly blown-up three-cycles.
In this framework, it has been shown~\cite{Cachazo:2001jy,cachazo} 
how to get the non-perturbatively generated effective superpotential 
of the dual \Ne{1} gauge theory~\cite{Taylor:1999ii,Vafa:2000wi} by means of
geometric considerations.

In this paper we bring together these lines of research, and show that
the explicit knowledge of the supergravity solution produced by a stack
of D-branes can be fruitfully combined with general geometric considerations
on the background in which they are embedded. In this way one can obtain
relevant information on the dual gauge theory, both at a perturbative and at a 
non-perturbative level. In particular we will discuss 
the Affleck--Dine--Seiberg theory, namely $U(N)$ \Ne{1} Super QCD with $N_f<N$ 
flavors of quarks and antiquarks, which we realize as the world-volume 
theory of a stack of fractional 
D3-branes on a \czz{} orbifold. We then show how to use this fractional brane 
configuration to obtain the running coupling constant, the 
classical moduli space of the low-energy theory and
the non-perturbative effective superpotential~\cite{Affleck:1983mk} of
SQCD in the chosen regime. We also comment on SQCD with $N=N_f$ which simply
arises as a particular case of our construction. 

This paper is organized as follows. In section~\ref{s:sqcd} 
we introduce the \czz{} orbifold and show how $U(N)$ SQCD can 
be engineered via a specific configuration of fractional D3-branes. 
After summarizing in section~\ref{s:sol} the corresponding supergravity 
solution, in section~\ref{s:moduli} we study the dual gauge theory, 
deriving the running gauge coupling constant 
and exploring the classical moduli space. 
Finally, in section~\ref{s:ads} we show how the non-perturbatively generated 
Affleck--Dine--Seiberg superpotential can be obtained by the fluxes of the 
dual supergravity solution together with some geometric considerations.

\vspace{0.5cm}
\section{SQCD from fractional branes on \czz}\label{s:sqcd}

The system  we are going to consider is a stack of fractional D3-branes
on the orbifold $\mathbb{R}^{1,3}\times\czz$~\cite{Douglas:1997de}.
In this space, which is a singular and non-compact Calabi--Yau
three-fold, a stack of D-branes will preserve four supercharges.
In particular we arrange the fractional branes according to Table~\ref{t:conf},
where $\,-\,$ and $\,\cdot\,$ 
indicate, respectively, longitudinal and transverse directions.
\begin{table}[h]
\begin{center}
\begin{tabular}{|c|c|c|c|c|c|c|c|c|c|c|}
\multicolumn{5}{c}{ }&
\multicolumn{6}{c}{$\overbrace{\phantom{\qquad\qquad\qquad\qquad\qquad}}^{\czz}$}\\
\hline
&0&1&2&3&\,4\,&\,5\,&\,6\,&\,7\,&\,8\,&\,9\,\\
\hline
D$3$ &$-$&$-$&$-$&$-$&$\cdot$&$\cdot$&$\cdot$&$\cdot$&$\cdot$&$\cdot$\\
\hline
\end{tabular}
\caption{Fractional D3-branes on \czz.\label{t:conf}}
\end{center}
\end{table}

In the following we will denote 
with $x^\alpha$ ($\alpha=0,\ldots,3$) the coordinates 
transverse to the orbifold, and introduce three complex 
coordinates in the orbifolded directions $x^r$ ($r=4,\ldots,9$) as follows
\begin{equation}\label{zi}
        z_1 = x^4 + i x^5~,\qquad
        z_2 = x^6 + i x^7~,\qquad
        z_3 = x^8 + i x^9~~.
\end{equation}

The generators of the two $\mathbb{Z}_2$ factors of the orbifold group 
are denoted by $g_1$ and $g_2$, and their action on the 
complex coordinates is given by
\begin{equation}\label{zzact}
\begin{tabular}{c|ccc}
        & $z_1$ & $z_2$ & $z_3$ \\
        \hline
        $g_1$ & $z_1$ & $-z_2$ & $-z_3$ \\
        $g_2$ & $-z_1$ & $z_2$ & $-z_3$ \\
\end{tabular}
\end{equation}
The remaining two elements of the orbifold group are of course 
the identity $e$ and $g_3 = g_1 g_2$.

\begin{figure}
\begin{center}
\includegraphics[scale=.5]{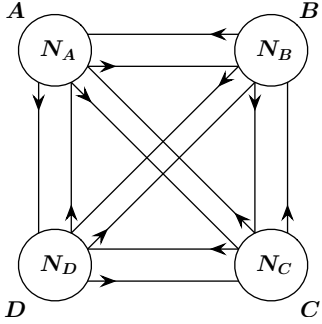}
\caption{{\small Each node $i$ of the quiver diagram represents a gauge group factor
$U(N_i)$ with the corresponding vector multiplet, and each oriented
arrow from node $i$ to node $j$ represents a chiral multiplet in the
$\hyp{N_i}{\bar{N}_j}$ representation.}}
\label{f:fullquiver}
\end{center}
\end{figure}

As is well known, the most elementary configurations of branes on orbifolds 
are made by fractional branes~\cite{Douglas:1996sw}.
These are defined by the fact that the Chan--Paton factors 
of the open strings attached to them transform in the irreducible 
representations of the orbifold group; moreover, they have
the property of being stuck at the orbifold fixed point. 
In our $\mathbb{Z}_2\times\mathbb{Z}_2$ orbifold, we then have four 
different types of fractional D3-branes, that we denote as A, B, 
C, D, corresponding
to the four irreducible one-dimensional representations of the orbifold group.
The low-energy theory living on a generic system of $N_k$ D3-branes of type 
$k$, is a four-dimensional \Ne{1} gauge theory with gauge group
$U(N_{\text{A}})\times U(N_{\text{B}})\times U(N_{\text{C}})
\times U(N_{\text{D}})$ and twelve chiral multiplets, which transform 
in the fundamental (or anti-fundamental) representation of a 
particular gauge group factor and carry a flavor index with respect
to the other three factors.
All this information can be encoded in the quiver diagram represented in
Fig.~\ref{f:fullquiver}.

\begin{figure}
\begin{center}
\includegraphics[scale=.6]{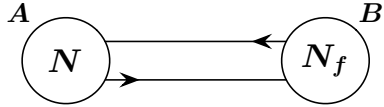}
\caption{{\small The quiver diagram associated to a system of $N$ fractional
D3-branes of type A and $N_f$ fractional D3-branes of type B which is used
to engineer $U(N)$ SQCD with $N_f$ flavors.}}
\label{f:SQCDquiver}
\end{center}
\end{figure}

Our goal is to use this orbifold set-up
to engineer $U(N)$ \Ne{1} SQCD, which is a theory with $N_f$ 
flavors of ``quarks'' and ``antiquarks''.
A possibility to do this would be to consider a stack of $N$
fractional D3-branes of a given type, and add to them $N_f$ 
D7-branes in order to introduce the fundamental and
anti-fundamental chiral multiplets~\cite{Marotta:2002gc}.
Here, however, we will follow an alternative and simpler 
route~\cite{Berenstein:2003fx}, which allows to obtain the main features of 
\Ne{1} SQCD in a very natural way by using only D3-branes.
This alternative strategy amounts simply to consider a 
configuration of $N$ fractional D3-branes of, say, type A and $N_f$ 
fractional D3-branes of, say, type B, which gives
rise to the $U(N)\times U(N_f)$ gauge
theory represented by the diagram of Fig.~\ref{f:SQCDquiver}.
From this quiver model we clearly can obtain $U(N)$ SQCD with $N_f$ flavors 
if we concentrate only on the low-energy degrees of 
freedom associated to the branes of type A by a suitable 
selection of open strings with appropriate Chan--Paton factors. 
Even if in the complete theory the flavor symmetry $U(N_f)$ 
is also gauged, we will see that this configuration of fractional 
D3 branes is able to provide a big deal 
of information on SQCD via the gauge/gravity correspondence.

\vspace{0.5cm}
\section{Summary of the supergravity solution}\label{s:sol}

The classical supergravity solution describing a system of $N_k$ fractional
D3-branes of type $k$ placed at the origin $z_1=z_2=z_3=0$ of \czz{}
was constructed in Ref.~\cite{Bertolini:2001gg}. Here we briefly summarize 
its main properties. 

Let us first recall that the bosonic fields of type IIB supergravity are
the metric, the 2-form $B_2$ and the dilaton $\phi$ from the NS--NS sector,
and the 0-form $C_0$, the 2-form $C_2$ and the 4-form $C_4$ with self-dual
field strength $\tilde{F}_5$ from the R--R sector. Then, the fractional
D3-brane solution is obtained by assuming that the two 2-forms $B_2$ and
$C_2$ have components only along the three exceptional shrinking 
2-cycles $\mathcal{C}_i$ 
($i=1,2,3$) of the orbifold geometry. Specifically, one writes 
\begin{equation}
        B_2 = b_i\ \omega_2^{(i)}~~, \qquad
        C_2 = c_i\ \omega_2^{(i)}~~,
\label{twistedfields}
\end{equation}
where the anti-self dual (1,1)-forms 
$\omega_2^{(i)}$, dual to the 2-cycles $\mathcal{C}_i$, 
are defined and normalized as
\begin{equation}\label{zzomeganorm}
        \int_{\mathcal{C}_i} \omega_2^{(j)} = \delta_i^j~~,\qquad
        \int \omega_2^{(i)} \wedge \omega_2^{(i)} = -\frac{1}{4}~~,\qquad
        \hd{4} \omega_2^{(i)} = - \omega_2^{(i)}~~.
\end{equation}
The scalar fields $b_i$ and $c_i$ in~\eqref{twistedfields} are called
twisted fields, and precisely correspond to massless degrees of
freedom of the three NS--NS and R--R twisted sectors of the conformal 
field theory
describing closed strings in the orbifold \czz{}.

If we define 
\begin{equation}
\label{G3}
G_3 = dC_2 + \big( C_0 + {\rm i}\,{\rm e}^{-\phi} \big) dB_2~~,
\end{equation}
and
\begin{equation}
\label{gammai}
\gamma_i = c_i +  {\rm i}\,b_i~~,
\end{equation}
then the solution for a stack of $N_k$ fractional D3 branes of type $k$ 
reads~\cite{Bertolini:2001gg}
\begin{equation}
\label{zzsol}
\begin{aligned}
        ds^2 &= H_3^{-1/2}\, \eta_{\alpha\beta}\, dx^\alpha dx^\beta\,
                +\,H_3^{1/2} \,\delta_{rs} \,dx^r dx^s~~,\\
        C_0&= 0~~, \qquad  {\rm e}^{\phi} ~=~ 1~~,\\
        \tilde{F}_5 &= d H_3^{-1}\, dx^0 \wedge \ldots \wedge dx^3
                + \hd{} \big( d H_3^{-1}\, dx^0 \wedge \ldots \wedge
          dx^3\big)~~,   \\
        G_3 &= d\gamma_i \wedge \omega_2^{(i)} ~~.
\end{aligned}
\end{equation}
Here the functions $\gamma_i$ are given by
\begin{equation}
\label{zzgamma}
        \gamma_i =  {\rm i} \,
K \,f_i(N_k) \,\ln \frac{z_i}{\epsilon_0}~~,
\end{equation}
where $\epsilon_0$ is a short-distance 
regulator, $K = 4\pi \gs \ls^2$ ($g_s$ being the
string coupling constant and $\ls$ the string length) and the $f_i$'s 
are functions of the numbers of the different types of fractional branes
\begin{equation}\label{fNk}
\begin{aligned}
f_1 (N_k) &= N_{\text{A}} + N_{\text{B}} - N_{\text{C}} - N_{\text{D}}~~,\\
f_2 (N_k) &= N_{\text{A}} - N_{\text{B}} + N_{\text{C}} - N_{\text{D}}~~,\\
f_3 (N_k) &= N_{\text{A}} - N_{\text{B}} - N_{\text{C}} + N_{\text{D}}~~.
\end{aligned}
\end{equation}
Finally, $H_3$ is a specific function of $z_i$ whose explicit expression, 
which we will not need in the following, can be found in 
Ref.~\cite{Bertolini:2001gg}.

As is clear from the solution~\eqref{zzsol}, each individual fractional 
D3-brane is charged under all three twisted sectors of the closed
string theory on the orbifold \czz{}, as well as under the untwisted one.
This is perfectly consistent with the description 
of fractional D-branes in terms of boundary states which represent
the sources for all closed string states emitted by the D-branes. In our
specific case, the boundary states for the various fractional D3-branes
are schematically given by
\begin{equation}
\label{bstates}
\begin{aligned}
\ket{\text{A}} &= \ket{\text{U}} + \ket{\text{T1}} + \ket{\text{T2}} 
+ \ket{\text{T3}}~~,\\
\ket{\text{B}} &= \ket{\text{U}} + \ket{\text{T1}} - \ket{\text{T2}} -
\ket{\text{T3}}~~,\\
\ket{\text{C}} &= \ket{\text{U}} - \ket{\text{T1}} + \ket{\text{T2}} -
\ket{\text{T3}}~~,\\
\ket{\text{D}} &= \ket{\text{U}} - \ket{\text{T1}} - \ket{\text{T2}} + 
\ket{\text{T3}}~~,
\end{aligned}
\end{equation}
where $\ket{U}$ is the contribution of the untwisted sector,
$\ket{\text{T}i}$ is the one of the $i$-th twisted sector
corresponding to the group element $g_i$, and the signs are consistent
with~\eqref{fNk}.

Finally, we remark that the supergravity solution~\eqref{zzsol} 
has a naked singularity of repulson
type, a common feature of all classical solutions describing fractional 
branes on orbifolds. One should then proceed to examine 
the appearance of an enhan\c con mechanism~\cite{Johnson:1999qt} that would 
make this geometry acceptable, but we will
not perform this analysis here. Instead, in the following sections
we will use the above solution to study the dual \Ne{1} gauge theory.

\vspace{0.5cm}
\section{The dual gauge theory and its classical moduli space}\label{s:moduli}

Let us now concentrate on the configuration made up of $N$ fractional 
D3-branes of type A and $N_f$ fractional D3-branes of type B that we 
introduced in section~\ref{s:sqcd} and represented in 
Fig.~\ref{f:SQCDquiver}. As we have seen, the theory 
on the world-volume of the type A branes is $U(N)$ SQCD with \Ne{1} 
supersymmetry, $N_f$ ``quark'' chiral multiplets, $Q^i$, and $N_f$ 
``antiquark'' chiral multiplets, $\tilde{Q}_{\tilde{\jmath}}$.

Many properties of this gauge theory can be explicitly 
obtained from the supergravity solution~\eqref{zzsol} 
(with $N_C=N_D=0$). As a first example, let us
consider the running gauge coupling constant 
$g_{\text{YM}}$, which, as shown in Ref.~\cite{Bertolini:2001gg}, can
be directly related to the twisted scalars $b_i$ of the dual 
geometry, according to
\begin{equation}
\frac{1}{\gym} = \frac{1}{8\pi\gs} \frac{1}{(2\pi\ls)^2}
\sum_{i=1}^{3} \,b_i~~.
\label{gym}
\end{equation}
The right hand side of this equation can be equivalently 
written also in terms of the flux of $G_3$ 
across an appropriate singular 3-cycle of the orbifold \czz{}.
To see this, and also for our later analysis of the superpotential, 
it is useful to identify the singular 3-cycles that exist 
in this non-compact Calabi--Yau space. 

\begin{figure}
\begin{center}
\includegraphics[scale=.5]{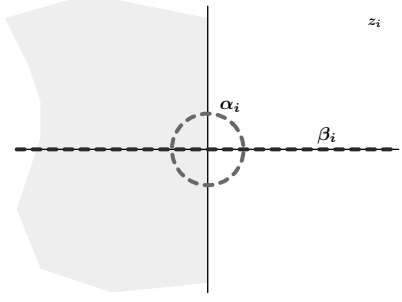}
\caption{{\small The compact 1-cycle $\alpha_i$ and the
noncompact 1-cycle $\beta_i$ in the $z_i$ plane.}}
\label{f:adscycles}
\end{center}
\end{figure}

Since there are no exceptional (1,2) or (2,1)-forms coming from the twisted sectors, everything
should arise from the (1,1)-forms $\omega_2^{(i)}$, dual to the singular
2-cycles $\mathcal{C}_i$ that we already encountered in section~\ref{s:sol}. 
Thus, we can introduce three compact 3-cycles $A_i$ and three 
non-compact 3-cycles $B_i$ by simply taking the direct product of
$\mathcal{C}_i$ with suitable 1-cycles in the $z_i$ planes. 
Specifically, we define
\begin{equation}\label{ABcycles}
        A_i = \alpha_i \times \mathcal{C}_i~~,\qquad
        B_i = \beta_i \times \mathcal{C}_i\qquad (i=1,2,3)~~,
\end{equation}
where the compact cycles $\alpha_i$ and the non-compact cycles $\beta_i$ in 
the $z_i$ plane are orthogonal to each other and are 
shown in Fig.~\ref{f:adscycles}.
Using the last equation of~\eqref{zzsol} and
the normalizations~\eqref{zzomeganorm}, one can easily see that 
the fluxes of $G_3$ along these 3-cycles are given by
\begin{equation}
\label{G3fluxes}
        \int_{A_i\,(\text{\,or }B_i)} \!G_3 = 
\int_{A_i\,(\text{\,or }B_i)} \!d\gamma_j \wedge \omega_2^{(j)}
                = \int_{\alpha_i\,(\text{\,or }\beta_i)} \!d\gamma_i ~
\int_{\mathcal{C}_i} \omega_2^{(i)}
                = \int_{\alpha_i\,(\text{\,or }\beta_i)}\! d\gamma_i~~.
\end{equation}
In particular, for our specific brane configuration, we find
\begin{equation}
\int_{A_1} \!G_3 = -2\pi K (N+N_f)~~,\qquad
\int_{A_2} \!G_3 = \int_{A_3} \!G_3 = -2\pi K (N-N_f)~~,
\label{afluxes}
\end{equation}
and
\begin{equation}
\int_{B_1} \!G_3 = {\rm i}\,K(N+N_f)\,\ln \frac{\rho_c}{\rho_0}~~,\qquad
\int_{B_2} \!G_3 = \int_{B_3} \!G_3 = 
{\rm i}\,K(N-N_f)\,\ln \frac{\rho_c}{\rho_0}~~.
\label{bfluxes}
\end{equation}
In the last line, the integrations over 
$\rho_i=\abs{z_i}$ in the non-compact cycles $\beta_i$ extend
up to a cut-off $\rho_c$, which sets the higher scale, 
while the lower scale 
$\rho_0$ is introduced as a further short-distance cut-off, 
since the validity of the singular 
supergravity solution stops at a finite distance from
the brane position.

If we identify $\rho_0$ with the (arbitrary) cut-off $\epsilon_0$
appearing in~\eqref{zzgamma}, we immediately see that $b_i = -{\rm i}
\int_{B_i} G_3$, and thus~\eqref{gym} becomes
\begin{equation}
        \frac{1}{\gym} = \frac{1}{8\pi\gs} \frac{1}{(2\pi\ls)^2}
                \sum_{i=1}^{3}\left(-{\rm i}\int_{B_i} G_3\right)
                = \frac{1}{8\pi^2} \big(3N - N_f\big) 
\ln \frac{\rho_c}{\rho_0}~~.
\label{running}
\end{equation}
Introducing the renormalization scale $\mu$ and the dynamically generated
scale $\Lambda$ through the usual energy/radius relations
\begin{equation}
        \rho_c = 2\pi\ls^2\ \mu~~,\qquad \rho_0 = 2\pi\ls^2\ \Lambda~~,
\end{equation}
that follow from a ``stretched string'' analysis~\cite{Peet:1998wn},
we easily get from~\eqref{running} the correct one-loop running coupling 
constant for our $U(N)$ SQCD theory.

We now turn to the vacuum structure of this theory and show how it
can be recovered from the supergravity solution~\eqref{zzsol} of
fractional D3-branes in a very simple way. 
Since it is known that 
$U(N)$ SQCD has a very different behavior depending on the number of flavors
(see for instance Ref.~\cite{Intriligator:1995au}), here 
we concentrate on the case in which the number of flavors 
$N_f$ is less than the number of colors $N$. This is the so-called 
Affleck--Dine--Seiberg (ADS) theory~\cite{Affleck:1983mk}, in which 
an effective superpotential is generated by non-perturbative effects.

To see how the vacuum structure of the ADS theory 
can be recovered from supergravity, 
let us begin by considering a single fractional D3-brane of type A 
together with one of type B. As individual objects, these branes are 
charged under all four sectors of closed string theory, and for this reason
they cannot move off the orbifold fixed point (see the corresponding
boundary states~\eqref{bstates}). 
However, since A and B are mutually BPS objects, we can easily construct
the superposition A$\,+\,$B, which will be charged under the sector 
twisted by $g_1$ (with a charge double with respect to a single 
fractional brane), but will not carry any charge 
under the other two twisted sectors. This can be easily seen in the 
boundary state description, where one writes
\begin{equation}
        \ket{\text{A}+\text{B}} = \ket{\text{A}} + \ket{\text{B}} = 2\,
\Big( \ket{\text{U}} + \ket{\text{T1}}\Big)~~.
\end{equation}
Thus, the superposition A$+$B can freely move in the $z_1$ plane,
which is left fixed by the action of $g_1$. Notice that when the pair A$+$B
is not at the origin, it cannot be divided anymore into its components,
which indeed are defined only at the orbifold fixed point. The crucial
observation is that the 
motion of A$+$B causes the breaking of the $U(1)\times U(1)$ gauge
group of the theory living 
on the superposition down to $U(1)$ via the Higgs mechanism.

\begin{figure}
\begin{center}
\includegraphics[scale=.6]{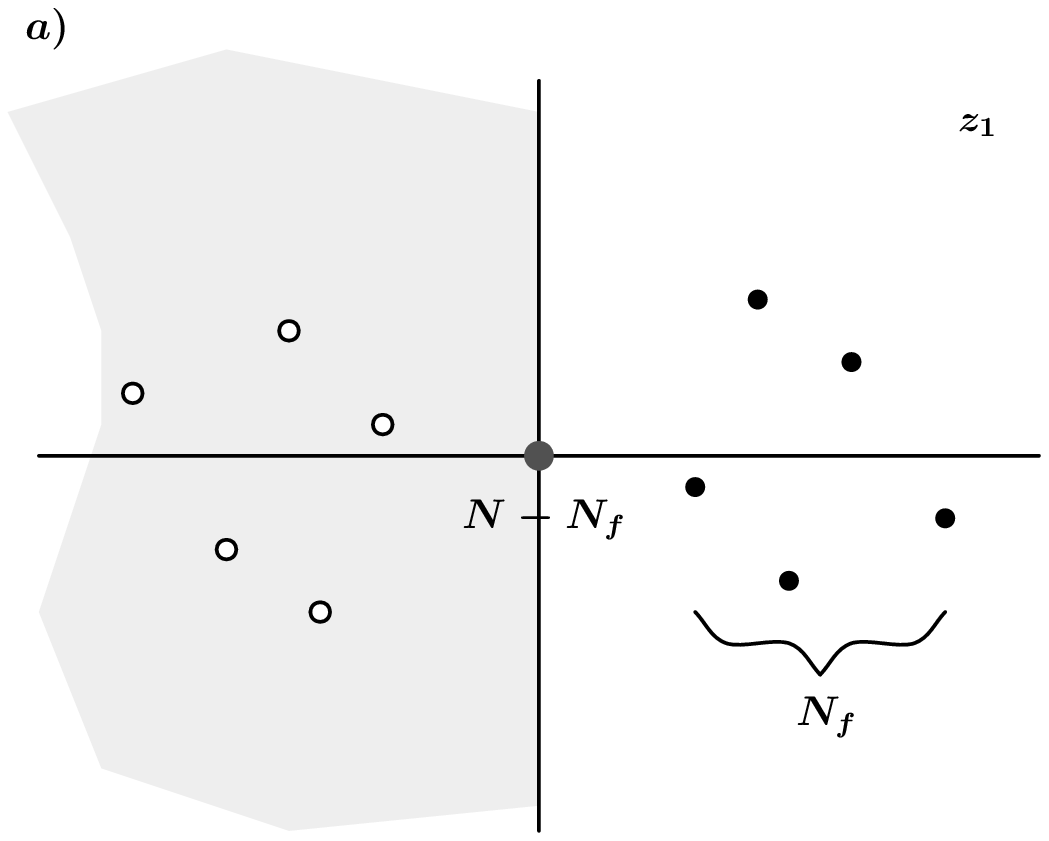}\qquad
\includegraphics[scale=.6]{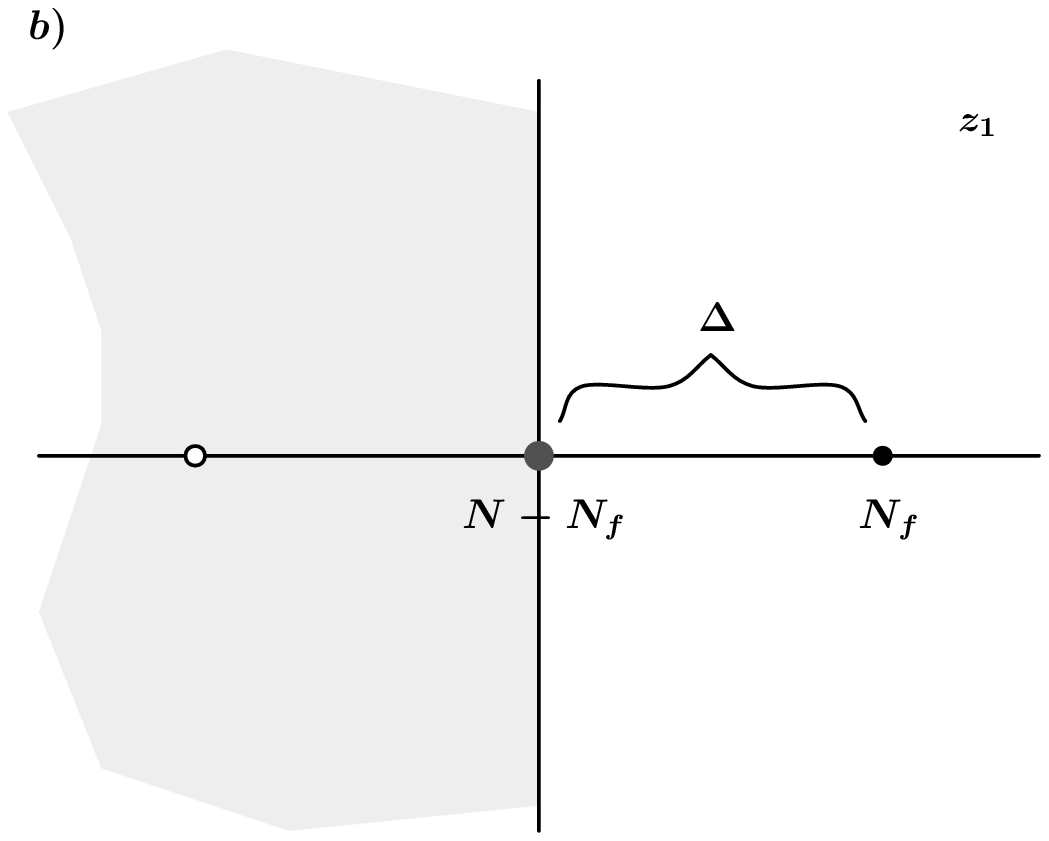}
\caption{{\small Moduli space of the ADS theory via fractional 
branes. \emph{a)} A+B superpositions at arbitrary points of the 
$z_1$ plane, together with their images on the covering space. \emph{b)} 
The configuration which makes the meson matrix $M^i_{\tilde{\jmath}}$ 
proportional to the identity.}}
\label{f:adsmoduli}
\end{center}
\end{figure}

Returning to our SQCD configuration and repeating the above argument, one finds
that out of the $N$ branes of type A and the $N_f$ branes of type B, 
it is possible to build $N_f$ A$\,+\,$B 
superpositions and move them away from the origin at arbitrary
points in the $z_1$ plane. The motion of these A$\,+\,$B pairs is naturally
interpreted as giving arbitrary vacuum expectation values to the meson
matrix $M^i_{\tilde{\jmath}} = Q^i \tilde{Q}_{\tilde{\jmath}}$, thus breaking
the gauge group $U(N)$ down to  $U(N-N_f)$ corresponding to the
world-volume theory of remaining $N-N_f$ fractional branes of 
type A still placed at the origin. Therefore, this D-brane construction 
uncovers the correct classical moduli space of the ADS theory 
in a very natural way, as shown in Fig.~\ref{f:adsmoduli}a.
This same mechanism works also for $N=N_f$. In this case, all branes can form
pairs and, since no unpaired A branes are left, no unbroken gauge theory 
remains. Notice that this description makes very clear that 
something drastic happens when passing from 
$N_f \le N$ to $N_f > N$, in agreement with the known field theory 
results~\cite{Intriligator:1995au}. Indeed, if $N_f > N$, all we have done 
is no longer valid and one is forced to look for some alternative description.

Let us now discuss how the supergravity solution~\eqref{zzsol} gets
modified when we form the A$\,+\,$B superpositions and move them in the 
$z_1$ plane. For simplicity, but without any loss of generality, 
we place all $N_f$ superpositions at the same point $z_1 = \Delta$ on the
real axis of the $z_1$ plane, as shown in Fig.~\ref{f:adsmoduli}b 
(clearly, in order to have an even configuration under the orbifold 
group, we also need to put images at 
the identified point $z_1 = -\Delta$). This set-up makes the meson 
matrix proportional to the identity, $M^i_{\tilde{\jmath}}
= m^2 \delta^i_{\tilde{\jmath}}$, where $m$ 
is related to $\Delta$ via the
usual gauge/gravity identification
\begin{equation}
\Delta = 2\pi\ls^2\, m~~.
\end{equation}
We will be interested in the case in which $\Delta$ and $m$ are
very large.

As we have seen before, what is relevant for the dual gauge theory is 
the knowledge of the twisted fields $\gamma_i$, which now become
\begin{equation}
\label{newgammai}
        \gamma_i = {\rm i} \,K \left[ (N - N_f) \ln \frac{z_i}{\epsilon_0}
                + \delta_{i,1}\ N_f\ \ln \frac{z_1-\Delta}{\epsilon_0}
                + \delta_{i,1}\ N_f\ \ln \frac{z_1+\Delta}{\epsilon_0} 
\right]~~.
\end{equation}
Then, using~\eqref{ABcycles} and~\eqref{G3fluxes}, we find that the flux
of $G_3$ along the $A_i$ cycles is given by
\begin{equation}
\int_{A_i} \!G_3 = -2\pi K (N-N_f)
\label{newafluxes}
\end{equation}
for all $i$'s, while the flux along the $B_i$ cycles becomes
\begin{equation}\label{gammab}
\int_{B_i} \!G_3    = {\rm i} \,K \left[( N- N_f ) \ln \frac{\rho_c}{\rho_0} 
+ \delta_{i,1}\ 2N_f\ \ln \frac{\Delta
}{\rho_0}\,\right]~~,
\end{equation}
where we have assumed $\rho_c \ll \Delta$ and denoted again by $\rho_0$ 
the minimum distance (from the branes at the origin 
and the pairs in $\pm \,\Delta$) that one can reach along the
integration path on $\rho_i=\abs{z_i}$ in the $\beta_i$ 
cycles.%
\footnote{Notice that if instead $\rho_c \gg \Delta$, the
flux of $G_3$ along the $B_i$ cycles is given by~\eqref{bfluxes}, in
agreement with the fact that in this case all degrees of freedom are light and
thus contribute to the running of the coupling constant as in~\eqref{running}.}

Thus, the gauge coupling constant for the world-volume theory
of the unpaired A branes is
\begin{equation}
\frac{1}{\gym} = \frac{1}{8\pi\gs} \frac{1}{(2\pi\ls)^2}
\sum_{i=1}^{3}\left(-{\rm i}\int_{B_i} G_3\right)
= \frac{3}{8\pi^2} (N - N_f) \ln \frac{\rho_c}{\rho_0}
+ \frac{2N_f}{8\pi^2} \ln \frac{\Delta}{\rho_0}~~,
\end{equation}
which in terms of the gauge theory scales becomes
\begin{equation}
\frac{1}{\gym} = \frac{3}{8\pi^2} (N - N_f) \ln \frac{\mu}{\Lambda}
+ \frac{2N_f}{8\pi^2} \ln \frac{m}{\Lambda}~~,
\end{equation}
with $\mu \ll m$. The above coupling can be expressed also 
in terms of the low-energy effective unbroken $U(N-N_f)$ theory as
\begin{equation}
\frac{1}{\gym}
= \frac{3}{8\pi^2}\, ( N - N_f )\, \ln \frac{\mu}{\Lambda_L}~~,
\end{equation}
where the low-energy scale $\Lambda_L$ is related to $\Lambda$ and $m$ via
\begin{equation}
\Lambda_L^{\phantom{L} 3 ( N - N_f ) }
= \frac{\Lambda^{ 3N - N_f }}{m^{2N_f}}
= \frac{\Lambda^{ 3N - N_f }}{\det M}~~.
\end{equation}
We then see that the supergravity computation precisely reproduces, 
beyond the running coupling constant, also the expected matching of scales 
in the gauge theory~\cite{Intriligator:1995au}. This concludes our 
analysis of the theory at the classical and perturbative levels. 
In the next section, we will turn to non-perturbative phenomena,
with the computation of the effective superpotential.

\vspace{0.5cm}
\section{The ADS superpotential}\label{s:ads}

We now study the quantum moduli space of SQCD using our D-brane
set-up, and see that our quantitative results perfectly agree with the field
theory analysis, predicting that the classical moduli space is lifted 
by the generation of the ADS superpotential~\cite{Affleck:1983mk}.

The main ingredient we will implement is the 
formula~\cite{Taylor:1999ii,Vafa:2000wi} that, in the cases where the dilaton
is constant, expresses the \Ne{1}
effective gauge superpotential $W_{\text{eff}}$ in terms of the fluxes of 
the complex three-form $G_3$ of the dual IIB supergravity solution 
and the periods of the holomorphic (3,0)-form $\Omega$ of the Calabi--Yau
orbifold in which the branes are put, namely
\begin{equation}\label{Vafa}
        W_{\text{eff}} ~\propto~ \sum_{i}\
                \left[\ \int_{A_i} G_3 \int_{B_i} \Omega
                - \int_{A_i} \Omega \int_{B_i} G_3\ \right]~~.
\end{equation}
As a brief summary, we recall that this formula has been 
shown to give the correct answer in many set-ups 
where the gauge theory is engineered via D-brane configurations on Calabi--Yau
manifolds which undergo a geometric 
transition, where some 3-cycles of the manifold blow up 
and the branes are replaced by fluxes of IIB 
supergravity fields~\cite{Gopakumar:1998ki,Vafa:2000wi}. For example, 
using this approach the Veneziano--Yankielowicz 
superpotential~\cite{Veneziano:1982ah}
of pure \Ne{1} super Yang--Mills theory has been 
extracted~\cite{Cachazo:2001jy,Giddings:2001yu} from the warped 
deformed conifold solution by Klebanov and Strassler~\cite{Klebanov:2000hb}.

In the case at hand, the supergravity solution we have at our 
disposal is not smooth, unlike the solution of Ref.~\cite{Klebanov:2000hb}. 
Rather, we are in a situation, the orbifold limit, where all 2- and 3-cycles 
are shrinking, similarly to what happens in the singular conifold 
solution of Klebanov and Tseytlin~\cite{Klebanov:2000nc}.
However, this does not seem to be an obstacle 
for using \eqref{Vafa}. In fact, in the
conifold case the knowledge of the singular
Klebanov--Tseytlin solution is sufficient for computing the $G_3$ fluxes
that are needed in \eqref{Vafa}, 
for the very simple reason that they are precisely 
identified with the fluxes of the regular Klebanov--Strassler 
solution.%
\footnote{This identification also explains the relation between the
radial coordinate $r$ appearing in the singular 
KT solution, and the coordinate $\tau$
appearing in the regular KS solution (see for example
Eq.~(100) in Ref.~\cite{Herzog:2002ih}).}
The same is true also in our case, and thus even if we do not
know a regular solution in a deformed orbifold, the $G_3$ fluxes can be
obtained from the singular solution as described in section~\ref{s:moduli}. 
On the other hand, the periods of $\Omega$ that enter in \eqref{Vafa}
cannot be determined in the singular case, since they crucially depend
on the details of the deformation of the geometry. However, the 
geometric considerations which are necessary for getting the correct 
periods do not depend on the details of the classical solution and can
be worked out in full generality also for the orbifold \czz{}.
Thus, we can fruitfully combine the knowledge of the fluxes coming from our
explicit supergravity solution with the geometric
features of the deformed background. This is the approach we take with our
fractional D-branes in the orbifold \czz{}.

We already derived the fluxes of $G_3$ along the $A_i$ and $B_i$ cycles: they
are given in~\eqref{newafluxes} and~\eqref{gammab}.
Let us then consider the periods of the holomorphic $(3,0)$-form $\Omega$. 
As in the case of the conifold, in order to get sensible results it is 
necessary to deform the singular geometry of the orbifold. 
Let us start by noting that the space \czz{} can be described as 
the $F(x,y,z,t)=0$ hypersurface in $\mathbb{C}^4$, where
\begin{equation}
        F(x,y,z,t) = xyz + t^2~~.
\end{equation}
The invariant variables in this function 
are related to the complex coordinates~\eqref{zi} by
\begin{equation}
        x = z_1^2~~,\qquad
        y = z_2^2~~,\qquad
        z = z_3^2~~,\qquad
        t =  {\rm i} \,z_1 z_2 z_3~~,
\end{equation}
and thus their engineering dimensions are $[x]=[y]=[z]=L^2$, and $[t]=L^3$. 
The simplest deformation of the complex structure, which also
resolves completely the singularity, is a constant deformation with parameter
$\xi$, namely
\begin{equation}
\label{deformation}
 F(x,y,z,t) ~~\to~~ F_\xi(x,y,z,t) = xyz + t^2 - \xi^2
\end{equation}
(notice that $[\xi]=L^3$). In Ref.~\cite{Berenstein:2003fx}, 
Berenstein has shown, via holomorphy considerations strengthened by 
a matrix model computation, that~\eqref{deformation} is indeed 
the correct deformation to consider. He also showed that the 
deformation parameter $\xi$ is related to the gaugino condensate $S$ 
of the dual gauge theory, as we too will argue below.
Given~\eqref{deformation}, we can write the holomorphic (3,0)-form
$\Omega$ in the usual way, according to
\begin{equation}
\label{Omega}
        \Omega = \frac{1}{2\pi {\rm i}} \oint_{F_\xi=0}
                \frac{dx \wedge dy \wedge dz \wedge dt}{F_\xi}
                ~=~ \frac{dx \wedge dy \wedge dz}
                {2\sqrt{\xi^2 - xyz}}~~.
\end{equation}
In order to compute the periods of $\Omega$ along a specific $A_i$ 
(or $B_i$) cycle, we define (with a little abuse of notation) 
$x=z_i^2$, $y=u+{\rm i}v$, $z=u-{\rm i}v$ and $\varepsilon^2 = 
\xi^2 / x$, so that from
\eqref{Omega} we get
\begin{equation}
\label{Omega1}
        \int \Omega = - {\rm i}\int \frac{dx \wedge du \wedge dv}
                {\sqrt{\xi^2 - x (u^2+v^2) }}
                = - {\rm i}\int \frac{dx}{\sqrt{x}} \int_{\mathcal{C}_i}
                \frac{du \wedge dv}{\sqrt{\varepsilon^2 - u^2 - v^2}}~~.
\end{equation}
The last integral can be easily evaluated, and is in fact a 
well-known result in the context of the geometry of the K3 manifold
\begin{equation*}
        \int_{\mathcal{C}_i}\frac{du \wedge dv}{\sqrt{\varepsilon^2 - 
u^2 - v^2}}
                = \int_{-\varepsilon}^{\varepsilon} du
                \int_{\gamma_u} \frac{dv}{\sqrt{\varepsilon^2-u^2-v^2}}
                = \int_{-\varepsilon}^{\varepsilon} du
                \int_{\gamma_\infty} \frac{dw}{{\rm i}\,w}
                = 4 \pi \varepsilon~~.
\end{equation*}
Using this inside~\eqref{Omega1}, we then have
\begin{equation}
        \int \Omega = -{\rm i}
\int \frac{dx}{\sqrt{x}} ~4\pi ~\frac{\xi}{\sqrt{x}}
                = - 4 \pi  {\rm i} \,\xi \int{\frac{dx}{x}}
                = - 8 \pi  {\rm i}\,\xi \int{\frac{dz_i}{z_i}}~~.
\end{equation}
Thus, the periods of $\Omega$ along the cycles $A_i$ and $B_i$ 
are finally given by
\begin{equation}
        \int_{A_i} \Omega = - 8 \pi  {\rm i} \,\xi \oint{\frac{dz_i}{z_i}}
                = 16 \pi^2 \xi~~,
\end{equation}
and
\begin{equation}
        \int_{B_i} \Omega
                = - 8 \pi {\rm i} \, 
\xi \int_{\xi^{1/3}}^{\rho_c} {\frac{d\rho_i}{\rho_i}}
                = \frac{8 \pi {\rm i}}{3}\ \xi\ \ln \frac{\xi}{\rho_c^3} ~~,
\end{equation}
where in the latter we have used the same upper cutoff $\rho_c$ already 
used in the computation of the fluxes of $G_3$, while 
the lower limit of integration must now be given by a suitable power of 
the deformation parameter $\xi$ (notice that since $\xi$ has dimension $L^3$,
this power is 1/3 in order to match with the length dimension of
$\rho_i=\abs{z_i}$).

We have now all the necessary ingredients to compute the 
effective superpotential of the gauge theory by means of
formula~\eqref{Vafa}. 
Inserting the appropriate dimensionful prefactors,
we find that the gauge effective superpotential is given by
\begin{equation}\label{weff}
\begin{split}
       W_{\text{eff}} &= \frac{1}{16\pi^2 {\rm i} \,K}
		\ \frac{1}{(2\pi\ls^2)^3}\ \sum_{i=1}^3\
                \left[\ \int_{A_i} G_3 \int_{B_i} \Omega
                - \int_{A_i} \Omega \int_{B_i} G_3\ \right]\\
	&= -\frac{1}{(2\pi\ls^2)^3}\ 
                \left[\ 3 ( N - N_f )\ \frac{\xi}{3}\ \ln \frac{\xi}{\rho_c^3}
                + 3( N - N_f )\ \xi\ \ln \frac{\rho_c}{\rho_0}
                + 2 N_f\ \xi\ \ln \frac{\Delta}{\rho_0}
                \ \right]~~.
\end{split}
\end{equation}
We now re-express the geometrical quantities in terms of gauge theory
quantities, by using again the ``stretched string'' energy/radius relation. 
Notice that the deformation parameter $\xi$, due to its engineering 
dimensions, is identified by the relation with a mass dimension 3 operator 
in the gauge theory, which is then natural to identify with the
gaugino condensate $S$ (see also Ref.~\cite{Berenstein:2003fx}). 
In summary the gauge/gravity relations are 
\begin{equation}
        \rho_c = 2\pi\ls^2\ \mu~~,\qquad
        \rho_0 = 2\pi\ls^2\ \Lambda~~,\qquad
        \Delta = 2\pi\ls^2\ m~~,\qquad
        \xi = (2\pi\ls^2)^3\ S~~,
\end{equation}
so that~\eqref{weff} becomes
\begin{equation}
        W_{\text{eff}} = - ( N - N_f )\ S\ \ln \frac{S}{\Lambda^3}
                - 2 N_f\ S\ \ln \frac{m}{\Lambda}~~.
\end{equation}
Though this result is correct, let us redefine the scales in 
order to write it in a more conventional way. The appropriate 
redefinition is $\Lambda\to e^{1/3}\Lambda$, $m\to e^{1/3}m$, and we get
\begin{equation}\label{VYT}
        W_{\text{eff}} = ( N - N_f )\ \left[\ S - S\ 
\ln \frac{S}{\Lambda^3}\ \right]
                - 2 N_f\ S\ \ln \frac{m}{\Lambda}~~,
\end{equation}
which is precisely the Taylor--Veneziano--Yankielowicz superpotential
for SQCD with $N_f$ flavors~\cite{Taylor:1983bp}. 
At the minimum the gaugino condensate is
\begin{equation}
        S = \left(\frac{\Lambda^{3N-N_f}}{m^{2N_f}}
\right)^{\frac{1}{N-N_f}}~~,
\end{equation}
so that from~\eqref{VYT} we get the ADS superpotential~\cite{Affleck:1983mk}:
\begin{equation}\label{ADS}
        W_{\text{eff}}  = ( N - N_f )\ \left[\
                \frac{\Lambda^{3N-N_f}}{m^{2N_f}}\ \right]
                ^{\frac{1}{N-N_f}}
                = ( N - N_f )\ \left[\
                \frac{\Lambda^{3N-N_f}}{\det M}\ \right]
                ^{\frac{1}{N-N_f}}\,.
\end{equation}
As an aside, we note that if $N_f = 0$ the above results 
reproduce the Veneziano--Yankielowicz superpotential 
for pure \Ne{1} Super Yang--Mills theory~\cite{Veneziano:1982ah}
\begin{equation}
W_{\text{VY}} = N \ \left[\ S - S\ \ln \frac{S}{\Lambda^3}\ \right]~~.
\end{equation}
Its value at the minimum (where $S=\Lambda^3$) is 
$W_{\text{VY}} = N \Lambda^3$. We stress that this is not 
just a formal limit of the result obtained for $N_f > 0$; indeed, one
could have started from the beginning by considering 
only $N$ fractional branes of type A, and apply formula~\eqref{Vafa}
to obtain the Veneziano--Yankielowicz superpotential.

Another observation concerns the case in which $N_f = N$. As we have mentioned
before, our brane construction can still be used in this case where all branes 
form A$\,+\,$B superpositions and no fractional branes (and thus no effective
gauge theory) are left. Therefore, the result for the moduli space 
can be read from the superpotential~\eqref{VYT} for $N_f = N$. In this case,
the minimization procedure implies $\det M = \Lambda^{2N}$, which is indeed 
the correct result expected for a \Ne{1} gauge theory with gauge group $U(N)$
and $N$ flavors.

Therefore we can conclude that our classical supergravity solution, 
together with some geometrical considerations, has been able to provide
relevant information on the \Ne{1} SQCD with $N_f$ flavors, both at
at the classical and perturbative level, and also at a non-perturbative
level. It would be very interesting to use this system of fractional 
branes to analyze SQCD also in the phase where $N_f > N$, where 
Seiberg duality is supposed to take place~\cite{Seiberg:1994pq}. 
Indeed a construction of Seiberg duality for generic quiver theories 
was presented in Refs.~\cite{Berenstein:2002fi,Berenstein:2003fx} 
with the implementation of quite formal methods involving also anti-branes. 
A more explicit analysis of these methods in the specific model we have
studied here is under current investigation~\cite{Imeroni:2003??}.

\subsection*{Acknowledgments}

We would like to thank Matteo Bertolini, Marco Bill\`o, 
Paolo Di Vecchia, Paolo Merlatti, Christian R\"omelsberger 
and Giuseppe Vallone for insightful discussions. E.I. also acknowledges 
feedback received at the Simons Workshop in Mathematics and Physics 
held at Stony Brook in August-September 2003 and at the RTN Workshop 
held in Copenhagen in September 2003. 
This work is partially supported by the European Commission 
RTN programme HPRN-CT-2000-00131, and by MIUR under contract
2001-1025492.

\addcontentsline{toc}{section}{References}
\providecommand{\href}[2]{#2}
\end{document}